\documentstyle[aps,12pt,epsf]{revtex}
\begin{document}
\tighten

\thispagestyle{empty}
\begin{flushright}
SLAC-PUB-95-7080\\
FERMILAB-PUB-95/401-T\\
December 1995\\
hep-ph/9601249
\end{flushright}
\vspace*{2cm}
\centerline{\Large\bf The $B_c$ Meson Lifetime}
\vspace*{1.5cm}
\centerline{{\sc Martin Beneke$^1$} and {\sc Gerhard Buchalla$^2$}}
\bigskip
\centerline{\sl $^1$SLAC Theory Group, Stanford University,}
\centerline{\sl P.O. Box 4349, Stanford, CA 94309, USA}
\vskip0.6truecm
\centerline{\sl $^2$Theoretical Physics Department}
\centerline{\sl Fermi National Accelerator Laboratory}
\centerline{\sl P.O. Box 500, Batavia, IL 60510, USA}

\vspace*{1.5cm}
\centerline{\bf Abstract}
\vspace*{0.5cm}
\noindent We investigate the total inclusive decay rate of 
the (ground state) $B_c$ meson
within the framework of an operator product expansion in 
inverse powers of the heavy
quark masses and subsequent matching onto nonrelativistic QCD.
The expansion is organized as a series in the strong coupling and 
in powers of the heavy quark velocities in the $B_c$,  
reflecting the nonrelativistic nature
of a heavy-heavy bound state. In this aspect the character of 
the expansion differs from the more familiar case of
heavy-light mesons. The framework incorporates systematically
corrections to the leading $b$- and $c$-quark decays due to binding
effects, as well as contributions from weak annihilation and
Pauli interference. Based on this approach we find for the
$B_c$ meson lifetime 
$\tau_{B_c}=(0.4 - 0.7)\,$ps, the dominant mechanism 
being the decay of the charm constituent.

\vspace*{1.5cm}

\vfill

\newpage
\pagenumbering{arabic}

\section{Introduction}
\label{intro}

The motivation for studying weak decays of
heavy hadrons is essentially twofold. First, one aims at 
understanding basic properties of the weak interaction at a 
fundamental level, including the precise determination of CKM
parameters. Second, systems containing heavy quarks allow us to test
our understanding of QCD in an interesting limiting case where, 
due to the large mass scale involved, certain aspects of the dynamics
simplify. Both topics are of course
intimately connected as the analysis of weak decays of heavy hadrons
is always faced with the problem of disentangling the interplay of
the strong and weak forces. In this respect the $B_c$ meson,
the lowest $\bar bc$ quark bound state, is a particularly interesting
system to study. Unlike in the case of heavy-light mesons, as
for example $B^+$, $B_d$ or $B_s$, the bound state dynamics of the
$B_c$ can be systematically treated in a nonrelativistic
expansion, which has proved very successful for the description
of the $c\bar c$ and $b\bar b$ family. At the same time, and by contrast
to the $c\bar c$ and $b\bar b$ ground states, $B_c$ is 
stable against strong or electromagnetic decay due to its 
flavor content and disintegrates only
via weak interactions. With these properties $B_c$ is in fact a unique
example, since the top quark lifetime is so short that the
analogous $t\bar b$ or $t\bar c$ mesons do not exist.

Several features of the $\bar bc$ system have already been the subject
of investigations in the past. A comprehensive analysis of the $\bar bc$
spectroscopy and the strong and electromagnetic decays of the excited
states has been given in \cite{EQ,Ger}. Weak decay properties of the 
groundstate $B_c$, semileptonic and various exclusive modes have also
been discussed \cite{DU,LM,CC}. The lifetime of $B_c$, $\tau_{B_c}$, is
briefly considered in \cite{LM,CC} where a rough estimate has
been presented. An estimate of $\tau_{B_c}$ using 
a modified spectator model and information
gained from the calculation of dominant exclusive modes is described in
\cite{QU}. The wide range of lifetimes, $\tau_{B_c}=(0.4 - 1.2)\,$ps, 
reported in these papers, reflects the various model assumptions on 
the modification of the free quark decay rates 
due to bound state effects.

In the present article we discuss a systematic approach to computing
the $B_c$ lifetime, which is
based on the optical theorem for the inclusive decay rate, an 
operator product expansion of the transition operator and a 
subsequent nonrelativistic expansion of the operator matrix elements 
in the $B_c$ meson state. In the first step, the operator product 
expansion of the transition operator, we rely on the fact, that if both 
quarks can be considered as heavy (and their mass difference is also 
large), the energy release in the weak decay of either quark is large
compared to the characteristic scale for the bound state dynamics. 
One may then expand in the ratio of these scales. Technically, this step 
copies the procedure for inclusive decays of heavy-light mesons, 
reviewed in \cite{BIG}. In the second step, the expansion of 
matrix elements, we utilize that the 
bound state of two heavy quarks is nonrelativistic in first approximation. 
This approximation can be systematically improved by means of 
nonrelativistic QCD \cite{CAS86}, which organizes the evaluation of 
matrix elements in full QCD in an expansion in $p/m_b$ and $p/m_c$, 
where $p= m_b v_b= m_c v_c\approx 1\,$GeV is the typical 
quark three momentum in the $B_c$ meson. The finite set of matrix elements 
in nonrelativistic QCD, which incorporates all nonperturbative effects, 
must be determined from lattice calculations or, less rigorously, from 
potential models. At this point, our treatment 
differs from an analogous one for heavy-light mesons, whose bound 
state dynamics is essentially different. In this case the scale  
relevant to the bound state is $\Lambda_{\mbox{\scriptsize QCD}}$, while in 
a heavy-heavy system an additional scale, $p>\Lambda_{\mbox{\scriptsize 
QCD}}$ (for the case at hand), is dynamically generated. Consequently, the 
importance of different operators is not ordered according to their 
dimension alone, but follows from the `velocity scaling rules' 
derived in \cite{LEP92}, when adapted to the case of two heavy quarks 
of different masses.

The procedure outlined above when combined with the inevitable emission 
of hard gluons in the decay, results in a double expansion in $\alpha_s$ 
and $v_Q\equiv p/m_Q$, with both 
parameters small, if both quarks are sufficiently 
heavy. 
Although this feature is conceptually very attractive, it is
not a priori clear whether in the realistic case the parameters
will be small enough to guarantee a reasonable behavior of the expansion. 
First, in the $c\to su\bar{d}$ transition, the ratio of the 
typical quark momentum in the bound state and the energy release is 
not a very small number. Second, although the 
reduced mass for the $\bar{b}c$ system falls in between the reduced 
masses of $\bar{b}b$ and $\bar{c}c$, the $c$ quark velocity in the 
$B_c$ meson is larger than in the $J/\Psi$, because the $c$ quark has 
to balance the momentum of a heavier $b$ quark. Thus relativistic 
corrections are expected to exceed those in the $\bar{c}c$ system. 
While it might appear
that the $B_c$ meson is a rather marginal case for the 
operator product expansion,
its convergence properties can only be properly assessed after an explicit
numerical investigation.
As we shall see, there are no obvious indications that the 
nonrelativistic expansion does not work.

We estimate a short $B_c$ lifetime (in comparison to earlier estimates), 

\begin{equation}
\tau_{B_c}=(0.4 - 0.7)\,\mbox{ps},
\end{equation}

\noindent the main uncertainty arising from the poorly known charm 
quark mass. 
$B_c$ mesons are expected to be produced at the LHC, if not before,
in numbers that are sufficient to test this prediction. Recently the 
first $B_c$ meson candidates have already been reported from LEP
\cite{LEP}. The search for $B_c$ at the Tevatron is summarized in 
\cite{tevatron}.

\section{The Operator Product Expansion}
\label{form}

The optical theorem relates the total decay width of a particle
to the imaginary part of its forward scattering amplitude. 
Applied to the $B_c$ meson total width $\Gamma_{B_c}$ this relationship
can be written as 

\begin{equation}
\label{gbc}
\Gamma_{B_c}=\frac{1}{2M_{B_c}}\langle B_c|{\cal T}|B_c\rangle
\end{equation}

\noindent where the transition operator ${\cal T}$ is defined by

\begin{equation}
\label{tdef}
{\cal T}=\mbox{Im}\ i\int d^4x \,T{\cal H}_{eff}(x){\cal H}_{eff}(0)
\end{equation}

\noindent 
Here ${\cal H}_{eff}$ is the usual effective hamiltonian describing
the low energy weak interactions of $b$ and $c$ quarks. The $B_c$ state
in (\ref{gbc}) is to be taken in conventional (relativistic) 
continuum normalization,
$\langle B_c|B_c\rangle=2 E V$.

In the case of heavy quark decay, where the energy release is large,
one may perform an operator product expansion (OPE) of ${\cal T}$.
In this way the expression in (\ref{tdef}) is expanded in a series
of local operators of increasing dimension, whose contributions to
$\Gamma$ are suppressed by increasing inverse powers of the heavy
quark masses. This formalism has already been applied to calculate
the total rates for charm and bottom hadrons containing one single
heavy quark \cite{BIG}. For earlier work using similar methods see
also \cite{SV,GRT}. Here we shall extend this approach to the 
treatment of $B_c$, where both constituents can be considered as heavy.
The operator product expansion leads to the generic result

\begin{equation}
{\cal T} = \sum_n C_n(\mu) {\cal O}_n(\mu).
\end{equation}

We now describe the operators that contribute to  
corrections up to order $v^4$ to the free quark decay, where $v$ is the 
quark velocity in the $B_c$ meson. For simplicity of discussion 
we will for a moment not distinguish the $b$ and $c$ quark 
velocities and masses (in 
the following $Q$ can be either $b$ or $c$). The velocity scaling 
rules themselves will be discussed in the subsequent section.
The dominant contribution to the decay width is generated by the 
operators ${\cal O}_{3Q}=\bar{Q} Q$. To leading order in the velocity 
expansion of its matrix element, this operator reproduces the free 
quark decay width of the quark $Q$. Using the equation of motion for 
the heavy quark fields to eliminate redundant operators, no operator 
of dimension four remains. There is one operator of dimension 
five, ${\cal O}_{GQ}=\bar{Q} g \sigma_{\mu\nu} G^{\mu\nu} Q$, whose 
contribution is suppressed by $v^4$ relative to the leading 
contribution. The most important correction comes from the 
dimension six operators ${\cal O}_{4Q}=\bar{b}\Gamma c\bar{c}
\Gamma' b$, which scale as $v^3$ ($\Gamma$, $\Gamma'$ 
collectively denote the Dirac and color structure). 
Finally, the dimension six operators
${\cal O}_{61Q}=\bar{Q}\sigma_{\mu\nu}\gamma_{\rho} D^\mu G^{\nu\rho} Q$ 
and ${\cal O}_{62Q}=\bar{Q} D_\mu G^{\mu\nu}\Gamma_\nu Q$ contribute 
also at order $v^4$. In the following we do not compute the 
coefficient functions of the latter two operators, 
so that the expansion will be complete to 
order $v^3$. We include the operators ${\cal O}_{GQ}$ and take their  
contribution as indicative of the error due to neglect of other 
contributions of order $v^4$.

Performing the OPE results in the expression,

\begin{equation}\label{tope}
{\cal T}={\cal T}_{35b}+{\cal T}_{35c}+{\cal T}_{6,PI}+{\cal T}_{6,WA},
\end{equation}
\begin{equation}\label{t6wa}
{\cal T}_{6,WA}={\cal T}_{cs}+{\cal T}_{ud}+\sum_l {\cal T}_{\nu l},
\end{equation}

\noindent where the first two terms in (\ref{tope}) account for the 
operators ${\cal O}_{3Q}$ and ${\cal O}_{GQ}$ and the other two for 
the four fermion operators ${\cal O}_{4Q}$. Explicitly,  

\begin{equation}\label{tb}
{\cal T}_{35b}=\Gamma_{b,spec} \bar bb-\frac{\Gamma_{0b}}{m^2_b}
\left[2 P_{c1}+P_{c\tau 1}+K_{0b}(P_{c1}+P_{cc1})+K_{2b}
(P_{c2}+P_{cc2})\right] {\cal O}_{Gb},
\end{equation}
\begin{equation}\label{tc}
{\cal T}_{35c}=\Gamma_{c,spec} \bar cc-\frac{\Gamma_{0c}}{m^2_c}
\left[(2 +K_{0c})P_{s1}+K_{2c}P_{s2}\right] {\cal O}_{Gc},
\end{equation}

\noindent where

\begin{equation}\label{g0b0c}
\Gamma_{0b}=\frac{G^2_Fm^5_b}{192\pi^3}|V_{cb}|^2\qquad\qquad
\Gamma_{0c}=\frac{G^2_Fm^5_c}{192\pi^3}
\end{equation}

\noindent and
\begin{equation}\label{k0k2}
K_{0Q}=c^2_-+2c^2_+ \qquad\qquad  K_{2Q}=2(c^2_+-c^2_-).
\end{equation}

\noindent The phase space factors $P_i$ are given as 
follows \cite{BIG,FLNN}:

\begin{equation}\label{pc12}
P_{c1}=(1-y)^4\qquad\qquad P_{c2}=(1-y)^3
\end{equation}
\begin{eqnarray}\label{pct1}
P_{c\tau 1}&=&\sqrt{1-2(r+y)+(r-y)^2}
  \left[1-3(r+y)+3(r^2+y^2)-
  r^3-y^3-4r y+7r y(r+y)\right] \nonumber \\
&& +\,12r^2 y^2\ln\frac{\left(1-r-y+\sqrt{1-2(r+y)+(r-y)^2}\right)^2}{4ry}
\end{eqnarray}
\begin{equation}\label{pcc1}
P_{cc1}=\sqrt{1-4y}(1-6y+2y^2+12y^3)+24 y^4 
   \ln\frac{1+\sqrt{1-4y}}{1-\sqrt{1-4y}}
\end{equation}
\begin{equation}\label{pcc2}
P_{cc2}=\sqrt{1-4y}(1+\frac{y}{2}+3y^2)-3y(1-2y^2) 
   \ln\frac{1+\sqrt{1-4y}}{1-\sqrt{1-4y}},
\end{equation}

\noindent where $y=m^2_c/m^2_b$ and 
$r=m^2_\tau/m^2_b$. $P_{s1}$ ($P_{s2}$) is
identical to $P_{c1}$ ($P_{c2}$), except that in this case
$y=m^2_s/m^2_c$. 
Note also that $P_{c\tau 1}=P_{c1}$ for $r=0$ and
$P_{c\tau 1}=P_{cc1}$ for $r=y$. In decays of the $b$ quark, we 
neglect terms of order $m_s^2/m_b^2$ and set $m_s=0$.
Furthermore

\begin{eqnarray}\label{tbcpi}
{\cal T}_{6,PI}&=&\frac{G^2_F}{4\pi}|V_{cb}|^2 p^2_-(1-z_-)^2\cdot \\
&\cdot& \left[
(c^2_++c^2_-)(\bar b_ib_i)_{V-A}(\bar c_jc_j)_{V-A}+
(c^2_+-c^2_-)(\bar b_ib_j)_{V-A}(\bar c_jc_i)_{V-A}\right] \nonumber
\end{eqnarray}
\begin{eqnarray}\label{tcs}
{\cal T}_{cs} &=& -\frac{G^2_F}{4\pi}|V_{cb}|^2 p^2_+\left[
\frac{(1-z_+)^3}{12}g_{\alpha\beta}+\left(\frac{(1-z_+)^2}{2}-
\frac{(1-z_+)^3}{3}\right)\frac{p_{+\alpha}p_{+\beta}}{p^2_+}
\right]\cdot \\
&\cdot&\left[
(c_+-c_-)^2(\bar b_ib_i)^\alpha_{V-A}(\bar c_jc_j)^\beta_{V-A}+
(5c^2_++c^2_-+6c_+c_-)(\bar b_ib_j)^\alpha_{V-A}
(\bar c_jc_i)^\beta_{V-A}\right] \nonumber
\end{eqnarray}
\begin{eqnarray}\label{tnut}
{\cal T}_{\nu\tau} &=& -\frac{G^2_F}{\pi}|V_{cb}|^2 p^2_+\cdot \\
&\cdot& \left[
\frac{(1-z_\tau)^3}{12}g_{\alpha\beta}+\left(\frac{(1-z_\tau)^2}{2}-
\frac{(1-z_\tau)^3}{3}\right)\frac{p_{+\alpha}p_{+\beta}}{p^2_+}
\right](\bar b_ib_j)^\alpha_{V-A}(\bar c_jc_i)^\beta_{V-A} \nonumber
\end{eqnarray}
\begin{equation}\label{trest}
{\cal T}_{ud}={\cal T}_{cs}(z_+\to 0)\qquad\qquad
{\cal T}_{\nu e}={\cal T}_{\nu\mu}={\cal T}_{\nu\tau}(z_\tau\to 0)
\end{equation}
\begin{equation}\label{pzdef}
p_\pm=p_b\pm p_c\qquad z_\pm=\frac{m^2_c}{p^2_\pm}\qquad
z_\tau=\frac{m^2_\tau}{p^2_+}
\end{equation}

\noindent In the equations above the QCD correction factors have
been written generically

\begin{equation}\label{cpm}
c_+=\left[\frac{\alpha_s(M_W)}{\alpha_s(\mu)}\right]^{6/(33-2f)}\qquad\qquad
c_-=\left[\frac{\alpha_s(M_W)}{\alpha_s(\mu)}\right]^{-12/(33-2f)}
\end{equation}

\noindent where $f$ is the number of flavors.
The scale $\mu$ is approximately $m_b$ in (\ref{tb}) and $m_c$ in (\ref{tc}),
respectively. For (\ref{tbcpi}) and (\ref{tcs}) one might anticipate 
a scale $\mu\approx 2 m_b m_c/(m_b+m_c)$, with twice the reduced mass
representing the characteristic scale of the $\bar bc$ bound state. 
Of course, the scales are not fixed precisely and one has to allow
for a certain variation in $\mu$, indicating an uncertainty that is
due to neglected higher order QCD effects. A clarification of
this issue requires the consideration of next-to-leading corrections.

\begin{figure}[h]
   \vspace{0cm}
   \epsfysize=7cm
   \epsfxsize=7cm
   \centerline{\epsffile{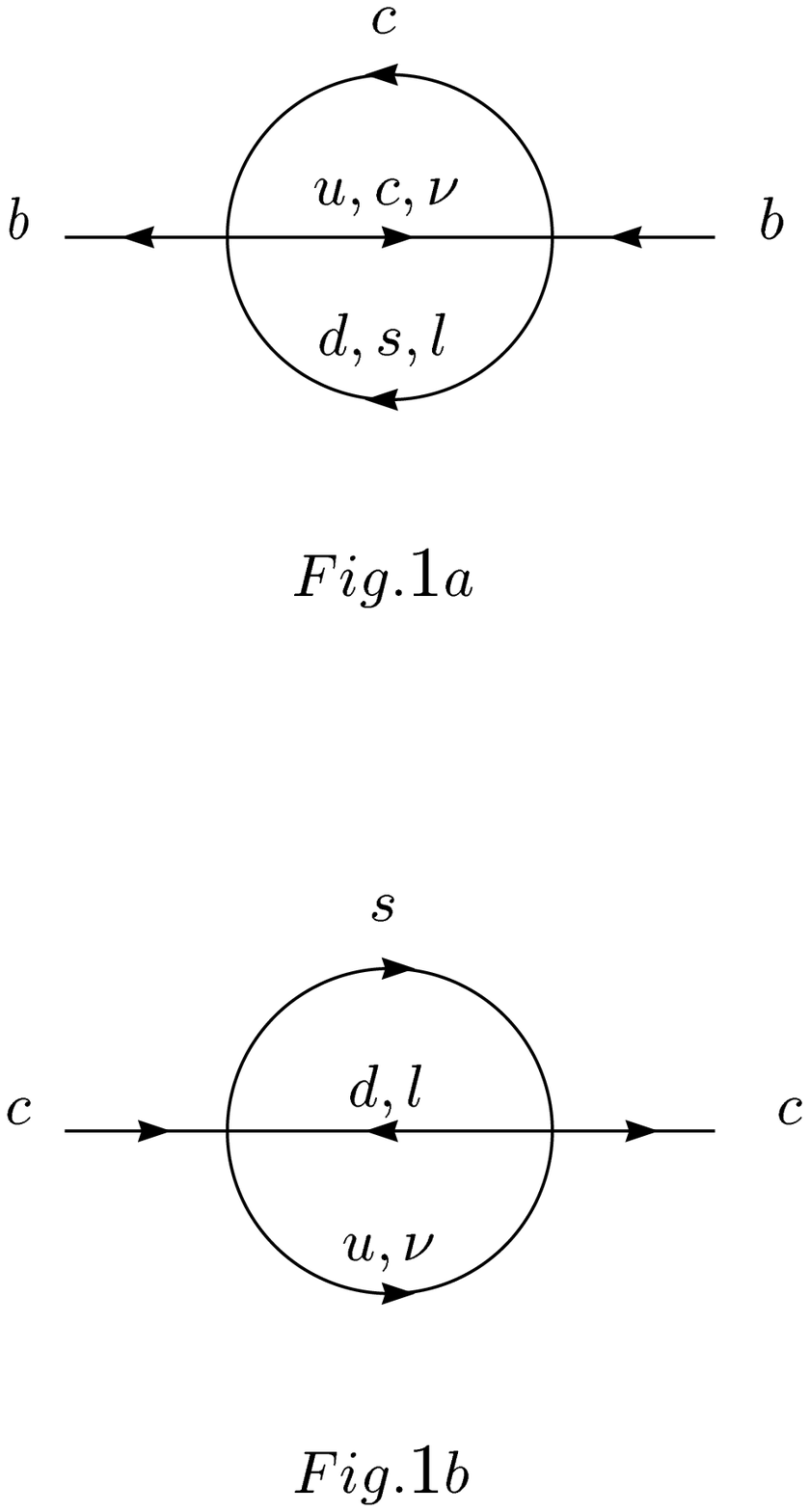}}
   \vspace*{1.2cm}
\caption{\label{bcfig1} Leading contributions in $\alpha_s$ to the 
spectator decays.}
\end{figure}

\begin{figure}[h]
   \vspace{0cm}
   \epsfysize=7cm
   \epsfxsize=7cm
   \centerline{\epsffile{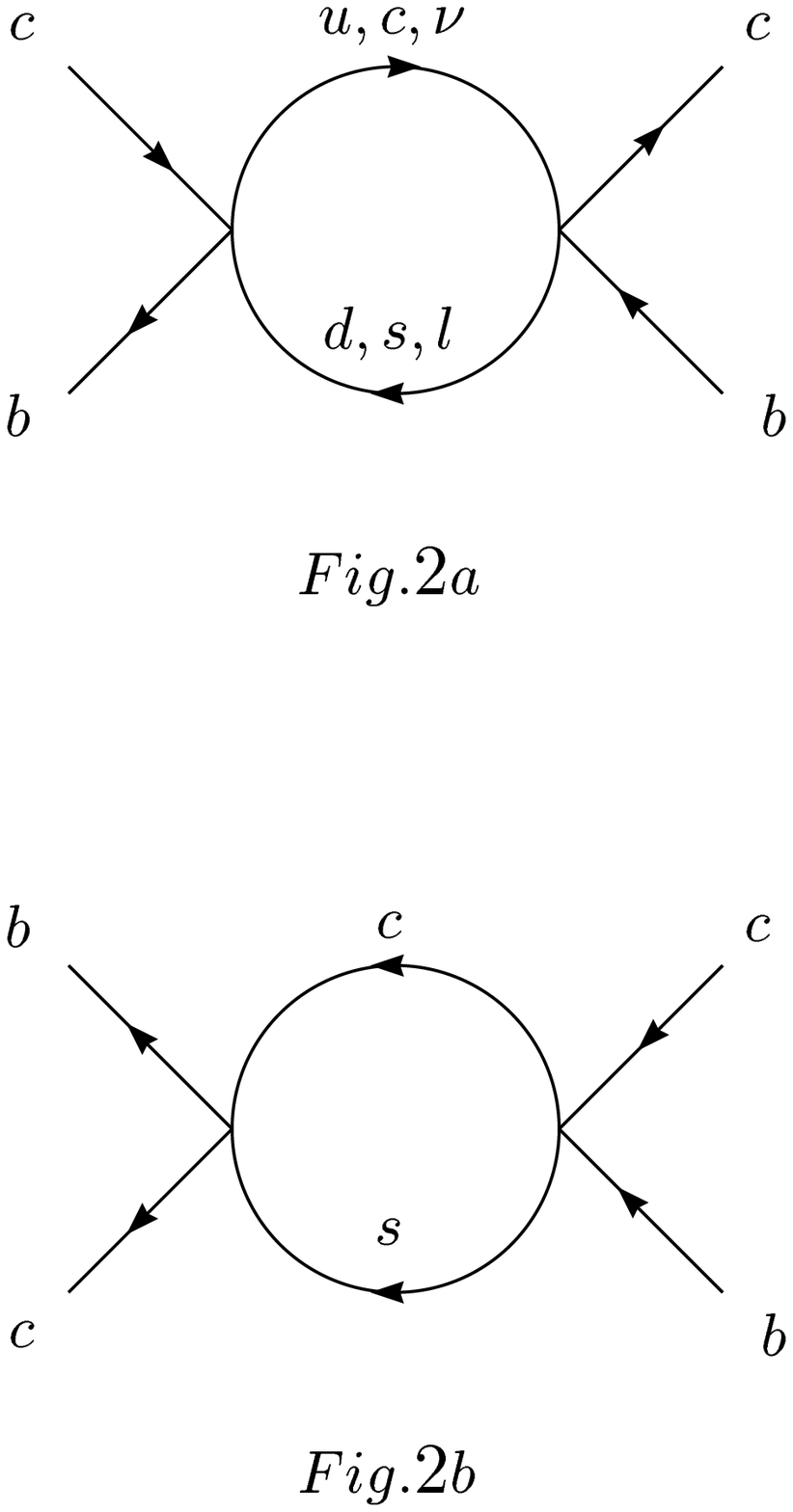}}
   \vspace*{1.2cm}
\caption{\label{bcfig2} `Weak annihilation' (a) and `Pauli interference' 
(b) contributions to the coefficient functions of four fermion operators 
in the OPE.}
\end{figure}

The contributions of the leading operators $\bar bb$ and $\bar cc$
correspond to the imaginary part of the diagrams in Fig.~\ref{bcfig1}, which
are contained in expression (\ref{tdef}).
The coefficients of $\bar bb$ and $\bar cc$ in (\ref{tb}), (\ref{tc})
can be obtained in the usual way by matching the diagrams 
of Fig.~\ref{bcfig1}, 
corresponding to the leading terms of the full expression (\ref{tdef}),
onto the operators $\bar bb$, $\bar cc$. These
coefficients are equivalent to the free quark decay rate and are 
known in next-to-leading logarithmic approximation in QCD
\cite{ACMP,BW,GB,HP,BBBFG} including the charm quark mass effects to 
${\cal O}(\alpha_s)$ \cite{BBBFG}. To include the next-to-leading log
effects the Wilson coefficients in the effective weak hamiltonians are
required at next-to-leading order and single gluon exchange
corrections to the diagrams in Fig.~\ref{bcfig1} need to be considered.
The complete next-to-leading order corrections are incorporated in the 
numerical analysis below and denoted by $\Gamma_{Q,spec}$ in 
(\ref{tb}) and (\ref{tc}). 
Similarly the contributions with ${\cal O}_{GQ}$ are obtained when an external
gluon line is attached in all possible ways to the inner quark lines in
Fig.~\ref{bcfig1}. The corresponding coefficients 
are known in leading log approximation.
Finally, the dimension six operators and their coefficients arise from
all those contributions, where one of the internal charm lines is 
`cut' in the diagrams of Fig.~\ref{bcfig1}a. The resulting graphs are
depicted in Fig.~\ref{bcfig2}. These contributions are also known 
in the literature as `weak annihilation' (Fig.~\ref{bcfig2}a) and 
`Pauli interference' (Fig.~\ref{bcfig2}b -- note the orientation of the 
fermion lines).

The expressions (\ref{tb}), (\ref{tc}) have been derived in
\cite{BBSUV} (see also \cite{MW}) and are also discussed in \cite{BIG}. 
The coefficients for the dimension six operators given in
(\ref{tbcpi})--(\ref{tnut}) are new. They are valid to leading
logarithmic accuracy in QCD and include charm quark and $\tau$ lepton
mass effects.

In order to obtain the total decay width, the expansion of the
transition operator (\ref{tope}) has to be taken between
the $B_c$ meson states. This step involves the calculation of
matrix elements of the local operators listed above. In general, this
is a difficult non-perturbative problem which in practice may
introduce considerable uncertainties in the evaluation of the
hadron lifetime. This is the case for instance for heavy-light
$b$ or $c$ flavored mesons \cite{BIG,BS}. By contrast, an important
advantage of the present system consisting of {\it two} heavy quarks
is the applicability of a nonrelativistic treatment. Given the
successes of this approach in describing $\bar cc$ and
$\bar bb$ quarkonia, one can expect a rather reliable determination
of the required $B_c$ matrix elements using the nonrelativistic
formalism, either within the framework of potential models or
by employing lattice QCD.
As we shall see, this treatment transforms the
$1/m_Q$ expansion into an expansion in powers of the heavy quark
velocities.

Before discussing the evaluation of matrix elements and the numerical
results for the $B_c$ meson lifetime we address briefly the
physical interpretation of the various terms in 
(\ref{tb}) -- (\ref{tbcpi}). First, let us consider the strictly
asymptotic limit in which $m_Q/\Lambda_{QCD}\to \infty$. In this
limit the $\bar bc$ system becomes extremely nonrelativistic,
consisting of two weakly bound heavy quarks, slowly moving in a
Coulomb type potential. The total decay rate for that system is then
simply given by the sum $\Gamma_{b,spec}+\Gamma_{c,spec}$ of the
weak decay rates of the quasi-freely and independently decaying
heavy quarks. In the asymptotic limit we have
$\langle B_c|\bar{Q} Q|B_c\rangle/(2M_{B_c})=1+{\cal O}(v^2_Q)$, 
$Q=b$, $c$,
and we indeed recover this simple picture within the formalism
described above. Of course, the truly asymptotic case is unrealistic.
However the simple spectator decay contribution is seen to emerge
as the formally leading term in the operator product expansion.
Bound state corrections are described in (\ref{tope}) by the
operators ${\cal O}_{GQ}$ (and others of higher dimension which we 
omitted) and corrections to the asymptotic value of the matrix 
element of $\bar{Q} Q$. 
Since the contributions
due to the chromomagnetic operators ${\cal O}_{GQ}$ can be related
to the $B_c-B^*_c$ mass splitting, it is actually of order $v^4$ 
in the nonrelativistic expansion and thus formally of sub-leading 
importance. The only correction of order $v^2$ arises from the 
kinetic energy of the quarks inside the bound state, which leads 
to a reduction of the matrix element of $\bar{Q} Q$ by a factor
$1-v_Q^2/2$, obviously representing the effect of time dilatation.

The question of bound-state corrections to the decay rate of a system
of two heavy fermions has also been studied, from a different point
of view (using a Bethe--Salpeter formalism), in \cite{KM} (see
also \cite{JTU}). It has been shown in this work, that the leading
net correction to the rate due to bound-state effects is just from
time dilatation, $-v^2/2$. Other conceivable
contributions, like phase space suppression and Coulomb enhancement
turn out to effectively cancel in the final answer. As we have seen,
this picture is in accordance with the findings obtained here within
the OPE approach, which incorporates this result in a rather
simple and straightforward manner.

\section{Matrix Elements}
\label{matel}

Let us next complete and summarize the discussion of the required
$B_c$ matrix elements, part of which we have already encountered
above. The operators described in the previous section 
are still expressed in terms of four-component 
fields $Q$. In a system containing a nonrelativistic quark, anti-quarks 
can not be produced, since this would require an energy larger than $m_Q$. 
It is then appropriate to `integrate out' the small components of the 
field and express the result in terms of a two-spinor $\psi_Q$. 
In this way, all contributions from scales larger than $\mu$, where 
$m_Q > \mu > m_Q v_Q$, are made explicit and can be accounted for 
perturbatively. This procedure is standard from nonrelativistic 
approximations to QED and Heavy Quark Effective Theory and leads to

\begin{eqnarray}
\label{barQQ}
\bar{Q} Q &=& \psi_Q^\dagger \psi_Q - 
\frac{1}{2 m_Q^2} \psi_Q^\dagger (i\vec{D})^2\psi_Q + 
\frac{3}{8 m_Q^4} \psi_Q^\dagger (i\vec{D})^4\psi_Q \nonumber\\
&&-\frac{1}{2 m_Q^2}\psi_Q^\dagger g \vec{\sigma}
\cdot\vec{B}\psi_Q-\frac{1}{4 m_Q^3}\psi_Q^\dagger (\vec{D}\cdot 
g\vec{E})\psi_Q + \ldots
\end{eqnarray}
\begin{equation}  
\label{sigmaB}
\bar{Q} g\sigma_{\mu\nu} G^{\mu\nu} Q = -2 \psi_Q^\dagger g\vec{\sigma}
\cdot\vec{B}\psi_Q - \frac{1}{m_Q} \psi_Q^\dagger (\vec{D}\cdot g\vec{E})
\psi_Q + \ldots,
\end{equation}

\noindent where we have already omitted the term 
$\psi_Q^\dagger \vec{\sigma}\cdot(g\vec{E}\times \vec{D})\psi_Q$ 
(spin-orbit coupling), whose 
matrix element vanishes in the valence Fock state of a pseudoscalar 
meson. 
The two-spinor $\psi_Q$ is here defined to have the same
normalization as $Q$,
\begin{equation}\label{psinorm}
\int d^3x\ \psi^\dagger_Q\psi_Q =\int d^3x\ Q^\dagger Q
\end{equation}
To the required order $\psi_Q$ is then related to the upper components
$\phi$ of $Q$
\begin{equation}\label{qphi}
Q\equiv e^{-imt}\left(\matrix{\phi\cr\chi\cr}\right)
\end{equation}
through
\begin{equation}\label{psiphi}
\psi_Q=\left(1+\frac{(i\vec{D})^2}{8m^2_Q}\right)\phi
\end{equation}
which can be verified by using the equations of motion.
Note also that the covariant derivative is understood to be in
the adjoint representation when acting on the chromoelectric field
\begin{equation}\label{dive}
(\vec D\cdot\vec E)=(\vec\partial T^a-g f^{abc}T^b\vec A^c)\vec E^a
\end{equation}
Equations (\ref{barQQ}) and (\ref{sigmaB}) are valid up to terms
of ${\cal O}(v^6)$. In the numerical analysis below we shall
also neglect the small term of ${\cal O}(\vec p^4/m^4_Q)$, which
yields the ${\cal O}(v^4)$ correction to time dilatation
\begin{equation}\label{tdil}
1-\frac{1}{2}\frac{\vec p^2}{m^2_Q}+
\frac{3}{8}\frac{\vec p^4}{m^4_Q}+\ldots =
1-\frac{1}{2}v^2_Q-\frac{1}{8}v^4_Q+\ldots =
\sqrt{1-v^2_Q}
\end{equation}
Radiative corrections modify the coefficients of the
chromomagnetic ($\vec\sigma\cdot\vec B$) and the `Darwin' 
($\vec D\cdot\vec E$) term in (\ref{barQQ}). In the present context
these effects can be neglected consistently.

The matrix elements of the operators on the right hand side of
(\ref{barQQ}) and (\ref{sigmaB}) can be evaluated in the 
nonrelativistic effective theory \cite{CAS86}, for example 
on the lattice, the obvious advantage being the 
possibility of using coarse 
lattices, since short-distance effects are already accounted for. To obtain 
a desired accuracy in $v^2$, the appropriate number of corrections to the 
leading order effective Lagrangian have to be retained. Preliminary 
studies of $B_c$ mesons on the lattice have recently appeared \cite{lattice}, 
but are not yet competitive with phenomenological potential models.

To assess the importance of various contributions, we recall \cite{LEP92} 
that the quark field $\psi_Q$ scales 
with the heavy quark three velocity as $(m_Q v_Q)^{3/2}$, a spatial 
derivative as $m_Q v_Q$, the electric field $gE$ as $m_Q^2 v_Q^3$ and the 
magnetic field $g B$ as $m_Q^2 v^4_Q$ (in Coulomb gauge). The coupling $g$ 
in a matrix element counts as $v_Q^{1/2}$. In the present case of a 
system with unequal quark masses, one has to keep in mind that additional 
factors from the mass ratio can enhance or suppress a given contribution. 
These scaling rules imply that the last term kept in (\ref{barQQ}) and 
(\ref{sigmaB}) is of the same order as the chromomagnetic term 
$\vec{\sigma}\cdot\vec{B}$, in contrast to the 
analysis of heavy-light mesons where this 
term is suppressed by $\Lambda_{\mbox{\scriptsize QCD}}/m_Q$ relative to 
the chromomagnetic interaction. The velocity counting relies on the 
inequality $p\sim 1\,\mbox{GeV} > \Lambda_{\mbox{\scriptsize QCD}}$, where 
$p$ is the typical quark momentum in the bound state.

Let us now turn to the evaluation of matrix elements in potential models 
for the bound state. Denote by $T= m_c v_c^2/2+m_b v_b^2/2$ the 
average total kinetic energy of the quarks in the bound state. Then, since 
$m_b v_b= m_c v_c$, we obtain\footnote{To avoid additional notation, the 
matrix elements are written for $b$ quarks rather than antiquarks.}

\begin{equation}
\frac{\langle B_c|\psi_b^\dagger (i\vec{D})^2 \psi_b|B_c\rangle}
{2 M_{B_c}\cdot m_b^2}\simeq v^2_b\simeq\frac{2 m_c}{m_b (m_c+m_b)}
\, T
\end{equation}
\begin{equation}
\frac{\langle B_c|\psi_c^\dagger (i\vec{D})^2 \psi_c|B_c\rangle}
{2 M_{B_c}\cdot m_c^2}\simeq v^2_c\simeq\frac{2 m_b}{m_c (m_c+m_b)}
\, T.
\end{equation}

\noindent The kinetic energy $T$ is known rather precisely, within 
about $10\%$,  
$T=0.37\,$GeV \cite{Ger}. It is also approximately independent of 
the reduced mass of the bound state (for the `logarithmic potential' 
this statement is exact). This value of $T$ implies 
$v_c^2\approx 0.38$ and $v_b^2\approx 0.035$. 
A straightforward calculation yields the 
matrix elements

\begin{equation}
\frac{\langle B_c|\psi_c^\dagger g \vec{\sigma}\cdot\vec{B}\psi_c
|B_c\rangle}{2 M_{B_c}} = -\frac{4}{3} g^2 \frac{|\Psi(0)|^2}{m_b}
\end{equation}
\begin{equation}
\frac{\langle B_c|\psi_c^\dagger (\vec{D}\cdot g\vec{E})\psi_c
|B_c\rangle}{2 M_{B_c}} = \frac{4}{3} g^2 |\Psi(0)|^2,
\end{equation}

\noindent where $\Psi(0)$ denotes the wavefunction at the origin. 
The matrix elements for $b$-quark fields are obtained by interchanging 
$m_b$ and $m_c$ on the right hand side. The second equation can also 
be obtained from using the equation of motion for the chromoelectric 
field. The matrix element of the resulting four fermion operator 
factorizes to leading order in $v^2$ and can be evaluated in a
straightforward way. 

We note that in the 
potential model $\Psi(0)$, the decay constant $f_{B_c}$ and the 
vector-pseudoscalar spin-splitting are related by

\begin{equation}
f_{B_c}^2=\frac{12 |\Psi(0)|^2}{M_{B_c}}\quad\quad
M_{B_c^*}-M_{B_c} = \frac{8}{9} g^2 \frac{|\Psi(0)|^2}{m_b m_c} .
\end{equation}

\noindent For the numerical analysis to follow we take 
$M_{B_c^*}-M_{B_c}=73\,$MeV from \cite{EQ}. With the parameters of 
the Buchm\"uller-Tye potential this spin-splitting implies 
$f_{B_c}=500\,$MeV, which we use as our central value for 
the decay constant.

Combining these 
results, we estimate the matrix elements (\ref{barQQ}) and 
(\ref{sigmaB}). Notice that the matrix element of 
$\bar{c}g\sigma_{\mu\nu} G^{\mu\nu}c$ is dominated by the divergence 
of the chromoelectric field rather than the spin-spin interaction, 
because the latter is suppressed by $m_c/m_b$. It is interesting 
to detail the deviations of the matrix element of $\bar{Q} Q$ from 
unity for the $c$-quark, where relativistic corrections are the 
largest. We find

\begin{eqnarray}
\frac{\langle B_c|\bar{c} c|B_c\rangle}{2 M_{B_c}} 
&=& 1- \frac{1}{2}v^2_c + \frac{3}{4}\frac{M_{B_c^*}-
M_{B_c}}{m_c}\left(1-\frac{m_b}{2 m_c}\right) + \ldots
\nonumber\\[0.1cm]
&\approx& 1-0.190+0.037-0.061+\ldots .
\end{eqnarray}

\noindent Despite the large $c$-quark velocity, this expansion 
appears quite well-behaved. As anticipated, the largest reduction 
of the rate is due to time-dilatation, representing the dominant 
bound state effect. Furthermore we have

\begin{equation}
\frac{\langle B_c|\bar{c}g\sigma_{\mu\nu} G^{\mu\nu}c|B_c\rangle}{2 M_{B_c}} 
= 3 m_c(M_{B_c^*}-M_{B_c})\left(1-\frac{m_b}{2 m_c}\right) 
\end{equation}

\noindent The matrix element of $\bar bb$ 
and $\bar{b}g\sigma_{\mu\nu}G^{\mu\nu}b$
can be obtained by interchanging $m_b\leftrightarrow m_c$
in the equations above.

Finally we need the matrix elements of the four quark operators in
(\ref{tbcpi}) -- (\ref{tnut}). To estimate their values, we 
employ factorization. This method, which lacks a firm footing for
heavy-light mesons, can in fact be justified in the present case of
$B_c$. Deviations from factorization arise from higher Fock 
components of the $B_c$ wavefunction and therefore are of higher order in
the nonrelativistic expansion. The relevant matrix elements are then
given by

\begin{equation}\label{me4ij}
\langle B_c|(\bar b_ib_j)^\alpha_{V-A}(\bar c_jc_i)^\beta_{V-A}
|B_c\rangle= f^2_{B_c}\left(\frac{1}{2}q^2 g^{\alpha\beta}-
q^\alpha q^\beta\right)
\end{equation}
\begin{equation}\label{me4ii}
\langle B_c|(\bar b_ib_i)^\alpha_{V-A}(\bar c_jc_j)^\beta_{V-A}
|B_c\rangle= \frac{f^2_{B_c}}{3}\left(\frac{1}{2}q^2 g^{\alpha\beta}-
q^\alpha q^\beta\right),
\end{equation}

\noindent where $q$ is the $B_c$ meson four-momentum, 
$q^2=M^2_{B_c}\approx (m_b+m_c)^2$, and $f_{B_c}$ is the $B_c$ meson
decay constant (in the normalization in which $f_\pi=131\,$ MeV). 
Deviations from factorization modify the decay rate only at order $v^5$ 
relative to the free quark decay. 
Using (\ref{me4ij}) and (\ref{me4ii}) the matrix elements of the
dimension six contributions to the transition operator are found to be

\begin{equation}\label{me6pi}
\frac{\langle B_c|{\cal T}_{6,PI}|B_c\rangle}{2 M_{B_c}}=
\frac{G^2_F}{12\pi}|V_{cb}|^2 f^2_{B_c} M_{B_c}p^2_-(1-z_-)^2
\left[ 2c^2_+-c^2_-\right]
\end{equation}
\begin{equation}\label{mecs}
\frac{\langle B_c|{\cal T}_{cs}|B_c\rangle}{2 M_{B_c}}=
\frac{G^2_F}{24\pi}|V_{cb}|^2 f^2_{B_c} M_{B_c} m^2_c (1-z_+)^2
\left[ 4c^2_++c^2_-+4c_+c_-\right]
\end{equation}
\begin{equation}\label{menut}
\frac{\langle B_c|{\cal T}_{\nu\tau}|B_c\rangle}{2 M_{B_c}}=
\frac{G^2_F}{8\pi}|V_{cb}|^2 f^2_{B_c} M_{B_c} m^2_\tau (1-z_\tau)^2
\end{equation}

\noindent The matrix elements of ${\cal T}_{ud}$, ${\cal T}_{\nu e}$ and
${\cal T}_{\nu\mu}$ are negligibly small as a consequence of
helicity suppression.

In the following section, we evaluate the QCD correction factors $c_{+/-}$ 
in the above expressions at a scale $\mu=2 m_{red}\approx 2.3\,$GeV, that 
is characteristic for a transition involving both bound state quarks. If 
this scale were widely separated from $m_b$, one would switch to the 
effective theory at a scale $\mu\approx m_b$, and 
scale the operators with 
their anomalous dimension to the lower scale. Since $2 m_{red}\approx m_b/2$, 
we excercise our freedom to choose the matching scale within some 
variation of $m_b$ and evaluate the QCD correction factors directly at 
$2 m_{red}$.

\section{Discussion}
\label{disc}

We are now ready to collect the various contributions presented
above and to derive an estimate for the $B_c$ meson lifetime. 
We first summarize the input parameters that we will use in our
numerical analysis:

\begin{equation}\label{mqnum}
m_b=5.0\,\mbox{GeV}\qquad m_c=1.5\,\mbox{GeV}\qquad m_s=0.2\,\mbox{GeV}
\qquad |V_{cb}|=0.04
\end{equation}
\begin{equation}\label{mmnum}
M_{B_c}=6.26\,\mbox{GeV}\qquad M_{B^*_c}-M_{B_c}=0.073\,\mbox{GeV}
\qquad T=0.37\,
\mbox{GeV}\qquad f_{B_c}=0.5\,\mbox{GeV}
\end{equation}

\noindent The parameters 
$M_{B_c}$, $M_{B^*_c}-M_{B_c}$ and $f_{B_c}$ are taken from \cite{EQ} 
and $T$ is taken from \cite{Ger}. 
The numbers for $m_b$ and $m_c$ correspond to pole masses, the
figures in (\ref{mqnum}) representing our central values. We 
comment on their choice below. The renormalization scales $\mu$ are 
chosen as follows: $\mu_1=m_b$ in decays of the $b$ constituent, 
$\mu_2=m_c$ in decays of the $c$ constituent and $\mu_3=2 m_{red}$ in 
modes involving both bound state quarks.

As already mentioned,
the pure spectator decay rates $\Gamma_{b,spec}$ and $\Gamma_{c,spec}$
are known to next-to-leading order in
perturbative QCD \cite{ACMP,BW,GB,HP,BBBFG}. The most complete
calculation, including final state mass effects in the QCD corrections,
is due to \cite{BBBFG}. We use their results to evaluate the 
spectator quark decay contributions. Thereby we take proper
account of charm quark and tau lepton mass effects. The strange quark
mass is kept for the charm decay modes, but neglected in $b$-decays.
The impact of electron and muon masses is likewise small and
has been neglected throughout.

First we would like to present an overview over the different
$B_c$ decay mechanisms and their relative importance as obtained
within the nonrelativistic OPE based framework we are advocating.
To this end we fix all input parameters at their central values
and calculate the partial, semi-inclusive $B_c$ decay modes
corresponding to the various underlying quark subprocesses.
The results are collected in Table I. We observe a dominance of
the charm decay modes over $b$-quark decay. Weak annihilation is
sizable, yet still considerably smaller than $b$-decay.
Adding up all contributions yields a total width of
$\Gamma_{B_c}=1.914\,\mbox{ps}^{-1}$, implying

\begin{equation}\label{taubc}
\tau_{B_c}=0.52\,\mbox{ps}
\end{equation}

\noindent Various branching fractions can also be inferred from Table I.
For instance the semileptonic branching ratio
$B(B_c\to Xe\nu)$ is found to be about $12\%$.

These estimates involve considerable uncertainties 
and Table I is merely intended to convey the general
trend and typical numbers for our favorite parameter set.
The dominant uncertainty comes from the quark masses. To limit
the related ambiguity in our phenomenological analysis we employ the
following strategy. We take the charm quark mass $m_c$ as the basic
input parameter, allowing it to vary within

\begin{equation}\label{mcrange}
1.4\,\mbox{GeV}\leq m_c\leq 1.6\,\mbox{GeV}
\end{equation}

\noindent which is in the ball park of 
the pole mass values available in
the literature. For any given value of $m_c$, we then fix the value
of $m_b$ by the requirement that the measured $B_d$ meson lifetime
$\tau_{B_d}\approx 1.55\,\mbox{ps}$ is obtained. This is justified
since the total $B_d$ width is essentially determined by the $b$-quark
decay contribution with very small pre-asymptotic bound state
corrections
and the OPE formalism can be expected to be reliable in this case.
For the $b$-quark spectator decay rate we again use the
complete next-to-leading order expressions \cite{BBBFG}. For
completeness we also incorporated the bound state effects that are
described in \cite{BIG}. Imposing the $\tau_{B_d}$ constraint
yields a $b$-quark mass that effectively includes unknown higher order
perturbative corrections in the decay of the bottom quark.
It practically
eliminates the $m_b$ and  $V_{cb}$ dependence of the predicted 
value of $\tau_{B_c}$. It turns out that a determination of $m_b$, 
for any given $m_c$,
via the approach just described, is approximately equivalent
to the relation

\begin{equation}\label{mbc35}
m_b=m_c+3.5\,\mbox{GeV}, 
\end{equation}

\noindent which is roughly consistent
with the well known formula relating $m_b$ and $m_c$ in HQET.
Adopting this procedure and varying $m_c$ between $1.4$ and 
$1.6$\,\mbox{GeV} results in the prediction

\begin{equation}\label{tbcrange}
0.4\,\mbox{ps}\leq \tau_{B_c}\leq 0.7\,\mbox{ps} .
\end{equation}

\noindent Variations of our central value (\ref{taubc}) due to 
variations of other input parameters can be inferred from 
Table~\ref{table2}. This table could also be used, if by the time 
$\tau_{B_c}$ is measured a different parameter set appears preferred.

In addition to using the measured $B_d$ lifetime one could be tempted 
to use the $D^0$ lifetime to eliminate the strong dependence on 
$m_c$. Since a very large charm mass is required to reproduce the absolute  
$D^0$ lifetime and semileptonic width in the OPE approach, this 
would lead to the prediction of a very low $B_c$ lifetime  
$\tau_{B_c}\approx 0.35\,$ps. We refrain from this procedure, since 
the OPE applied to charmed mesons is less reliable than for $B_c$ 
and probably more qualitative than quantitative \cite{BS}.

Let us conclude the discussion with the following remarks.

(a) As already discussed above, the expression for the $B_c$ meson
total width we have derived is consistent up to and including terms
of ${\cal O}(v^3)$ in the nonrelativistic
expansion. Part of the ${\cal O}(v^4)$ corrections, those due to the
operators $\bar Q g \sigma_{\mu\nu}G^{\mu\nu}Q$, have also been
considered to obtain an estimate of the order of magnitude of
these contributions, while the neglected contributions arise 
from the operators ${\cal O}_{61Q}$ and ${\cal O}_{62Q}$ listed 
earlier. The labor involved in calculating these remaining  
contributions does not appear justified in view of the uncertainties 
connected with the quark masses. To reinforce this point and 
to see the behaviour of the expansion more
explicitly, we write down the size of the various terms of a given 
order in $v$, using central parameter values.
For the sum of the $c\to s$ decay modes one obtains
$(1.526-0.289-0.008)\,$ps$^{-1}$,
writing separately the terms of order $v^0$,
$v^2$ and $v^4$. For the total $b$-decay
contribution one finds $(0.640-0.011-0.014)\,$ps$^{-1}$.
The sum of all $v^3$ effects amounts to typically $0.1\,$ps$^{-1}$.
In the case of $b$-decay the corrections are particularly small,
since they are additionally suppressed by 
inverse powers of the large $b$-quark mass.
Generally speaking, the relevant velocity is the one of the lighter
constituent, $v_c$, which determines the convergence properties 
of the nonrelativistic expansion. As we see, the series is rather
well behaved. Note also that a further contribution
of ${\cal O}(v^4)$ comes from the expansion of the time dilatation
factor. Numerically, $-v^4_c/8\approx -2\%$.
Although one has to be careful about drawing definitive conclusions,
since the ${\cal O}(v^4)$ contribution has not yet been
calculated completely, these observations support 
the assumption that the total ${\cal O}(v^4)$ term is rather small.

Let us re-emphasize that, technically, in the 
approach adopted here the velocity expansion 
distinguishes a heavy-heavy meson from a heavy-light meson, 
for which 
an expansion in $\Lambda_{\mbox{\scriptsize QCD}}/m_Q$ is appropriate.
This distinction was missed in the treatment of \cite{BIGBC}  
(which also starts with an OPE), 
resulting in an incorrect evaluation of the matrix elements
of $\bar Q g\sigma\cdot G Q$. As a consequence a large correction
was obtained from this formally subleading (${\cal O}(v^4)$)
contribution, which we found to be essentially negligible, even
for charm. The leading correction terms to the spectator picture
of ${\cal O}(v^2)$ from the kinetic energy and of
${\cal O}(v^3)$, from weak annihilation and Pauli
interference, were not calculated explicitly in \cite{BIGBC}.

(b) The Pauli interference contribution 
exhibits a fairly substantial
dependence on the renormalization scale. 
The number shown in Table~\ref{table1} corresponds to 
$\mu=2 m_{red}\approx 2.3\,\mbox{GeV}$. Allowing a variation of
$\mu$ from $1-5\,$GeV yields a range of values from 
$-0.342\,$ps$^{-1}$ to $-0.036\,$ps$^{-1}$ for this contribution.
This large dependence is formally of
${\cal O}(\alpha_s)$, which goes beyond the leading log approximation
we are working in presently. It represents a theoretical uncertainty
in this calculation. The pronounced sensitivity to the scale arises
from sizable cancellations that occur between the Wilson
coefficients in (\ref{me6pi}). Note that the Pauli interference
contribution is {\it positive\/} in the limit $\alpha_s\to 0$ and
changes sign due to the presence of important short-distance
QCD effects. The situation could be improved by studying
next-to-leading order QCD corrections to Pauli interference, including
a proper matching of the operators to non-relativistic QCD. 
By contrast, the scale dependence turns out to be very moderate
in the case of weak annihilation of $\bar bc$ into $c\bar s$,
only $\pm 10\%$ for the same range of $\mu$ as before.

(c) For the channels $\bar b\to\bar cc\bar s$ and
$\bar b\to\bar c\tau\nu$ the weak annihilation term can be
comparable to, or even exceed
the spectator contribution. This feature is related to the
strong phase space suppression of the three body decay mechanism,
which is absent for weak annihilation. It would persist even in the
limiting case where $\alpha_s$ is taken to be very small and the
nonrelativistic bound state description would be a perfect
approximation. The charm mass could be so large that the spectator
decay would be kinematically forbidden in which case weak  
annihilation would just be the leading contribution, but still reliably
calculable in the nonrelativistic approach.
For these reasons the dominance of weak annihilation does not  
indicate a problem for the validity of the operator product  
expansion.

(d) An important conceptual point that needs to be mentioned is that the
application of the OPE is based on the assumption of local quark hadron
duality. By this we mean that the sum over all decay channels can be 
described in terms of partonic degrees of freedom.
Although this assumption should be valid asymptotically
as $m_Q/\Lambda_{QCD}\to\infty$, it is not a priori clear how well
it is satisfied in practice. 
Little is known rigorously and quantitatively about the exact
conditions to be met if duality is supposed to hold. Conceivably, 
the issue of duality is related to the fact that the velocity 
expansion is at best asymptotic.
This includes the possibility that it might
not even be asymptotic at minkowskian kinematics. Since, for $B_c$, the 
expansion is still convergent, we are concerned mainly with 
the second possibility, in which case the OPE is meaningless 
beyond a certain order, even if it appears to be convergent. 
Although duality has been questioned for the decay of $D$ mesons 
\cite{BDS}, it seems, at least to the authors, that there is as yet no 
place in the charm and bottom hadron family 
where this possibility could be unambiguously separated 
from uncertainties due to unknown input parameters (quark masses, 
matrix elements) or convergence properties of the expansion.
In this situation one has to rely on a more pragmatic attitude
towards the problem, assume that quark-hadron duality makes sense
for a given case, explore the consequences and, if possible, check
whether the emerging picture is at least consistent.

The critical point with regard to this issue for $B_c$ is charm decay.
Here the energy release is not as comfortably large as it is in
the case of bottom decay, and in fact slightly smaller than in $D$ 
meson decay. However, 
the reliability of a duality is channel dependent. 
Duality might still
work reasonably well in the case of the charm transitions
contributing to $B_c$ decay, even if it were violated for
$D$ mesons. After all, as we have emphasized, the nature of the OPE
is very different for $B_c$ due to the nonrelativistic character
of the heavy-heavy type bound state. 

A useful cross-check comes from comparing the inclusive decay width 
for a given channel with the sum of the corresponding exclusive 
decay modes. In the present case, $c\to s u\bar{d}$,  these are modes like
$B_c\to B^{(*)}_s\pi$, $B^{(*)}_s\rho$, $B^{(*)}K^{(*)}$,
$\ldots$. In \cite{LM,CC} phenomenological models have been used to 
estimate the rates for twenty of these two-body decay modes.
Adding their results one finds a total of $0.49\,$ps$^{-1}$ (BSW model)
and $0.65\,$ps$^{-1}$ (ISGW model) in \cite{LM} and $0.67\,$ps$^{-1}$  
in \cite{CC}, to be compared with 0.91 ps$^{-1}$ for the inclusive width 
from Table~\ref{table1}. Viewing this comparison with due caution,
regarding the model dependences and other uncertainties in the 
estimation of exclusive modes as well as the charm mass uncertainty 
for the inclusive prediction, it is
reassuring that, first, the order of magnitude comes out to
be consistent and, second, the sum of exclusive modes does not
exceed the inclusive result. A similar observation holds for the
semileptonic $c\to se\nu$ transitions.

Although the arguments we have given are of a somewhat heuristic
nature, they all seem to indicate that the OPE approach makes
sense for inclusive $B_c$ decays. At the very least the underlying
assumptions do not seem to be obviously violated.

(e) In \cite{LM,QU} the $B_c$ lifetime has been estimated on the basis
of a modified spectator model, where the phase space for the 
free quark decay is modified to account for the physically accessible 
kinematic region \cite{LM} or 
the $b$ and $c$ quark masses
are reduced by the binding energy to incorporate bound state effects 
\cite{QU}.
Like the OPE formalism these approaches lead to a reduction of
the free quark decay rates caused by binding, but the details of
the resulting picture are markedly different. The modified spectator 
ansatz does not correctly approach the asymptotic limit of very 
heavy quark masses, where, as in the OPE-based calculation, 
the leading bound state corrections come only from time dilatation. 
Our bound state corrections are 
numerically much smaller than the very large effects reported in
\cite{QU}, which invert the hierarchy between $c$ and $\bar b$ decays,
making the latter more prominent, and lead to a considerably
longer lifetime of $\tau_{B_c}=1.35\pm 0.15\,$ps \cite{QU}.
It has been emphasized in \cite{QU} that the model calculation has to
be regarded with caution, since the inclusive charm decay rate is
found to be below the rate estimates of the corresponding exclusive modes. 
Due to this feature the result for the lifetime quoted above has been
corrected to $\tau_{B_c}=(1.1 -1.2)\,$ps \cite{QU}.
This estimate is however still sizably larger than the one derived in
the present analysis, where, as we have seen, the inclusive
and the exclusive approach can be viewed as roughly consistent.

\section{Summary}
\label{summ}

In this article we have presented a detailed investigation
of the $B_c$ meson total width based on a systematic
operator product expansion of the transition operator.
In several important aspects this analysis goes beyond the 
estimates derived previously in the literature:
First, we have emphasized the nonrelativistic nature of the
heavy-heavy bound state system $B_c$. As we have shown, this fact
has crucial implications for the organization of the operator
product expansion, which is different from the case of
heavy-light mesons. In addition the framework allows a systematic
evaluation of the relevant matrix elements of local operators using
the nonrelativistic bound state wave functions. Matrix elements
of four-quark operators, for instance, factorize 
exactly to leading order in the velocity expansion in contrast to the
general situation.

Free-quark decay of $\bar b$ and $c$ represents, both formally
and numerically, the leading (in $v$) contribution to the
$B_c$ decay rate. For their evaluation the complete
next-to-leading order expressions in renormalization group
improved QCD perturbation theory have been employed.
The dominant bound state corrections come from time dilatation and
are determined by the heavy quark velocities, which we have estimated
from existing potential model calculations. A further new result are
the complete ${\cal O}(v^3)$ contributions to the decay rate, due to
weak annihilation and Pauli interference, for which explicit expressions,
valid to leading logarithmic accuracy in QCD, have been derived.

Numerically our analysis yields $\tau_{B_c}=0.4 - 0.7\,$ps,
where the measured $B_d$ lifetime has been used to reduce the 
uncertainties due to the bottom quark mass and $V_{cb}$.

We have also briefly considered the reliability of the operator 
product expansion, whose applicability relies on the assumption of
quark-hadron duality. Although this issue remains an important
caveat to be kept in mind, we have found no indication for an 
obvious violation of its validity. On the contrary, the series
itself is rather well behaved and model calculations of
exclusive modes are consistent with the duality assumption.
It will be interesting to compare the predictions for $B_c$ decay based
on local duality with future experimental results. These should help
to assess to what extent the underlying theoretical assumptions
are valid in practice. Clearly, experimental progress towards a
measurement of the lifetime and other inclusive properties of $B_c$
is eagerly awaited.

\section*{Acknowledgements}

We thank Isi Dunietz for discussions and encouragement, 
Arthur Hebecker for discussions and
Estia Eichten and Chris Quigg for interesting discussions
and a critical reading of the manuscript.
The hospitality of the Aspen Center for Physics, where this study
has been initiated, is also gratefully acknowledged. 
Fermilab is operated by Universities Research Association, Inc.,
under contract DE-AC02-76CHO3000 with the United States Department
of Energy.

\vfill\eject

\newpage
\begin{table}[h]
\addtolength{\arraycolsep}{0.2cm}
\renewcommand{\arraystretch}{1.3}
$$
\begin{array}{|c|c|}
\hline
\mbox{Mode} & \mbox{Partial Rate in ps}^{-1} \\ 
\hline\hline
\bar{b}\to \bar{c} u\bar{d}        & 0.310 \\ \hline
\bar{b}\to\bar{c} c \bar{s}        & 0.137 \\ \hline 
\bar{b}\to\bar{c} e \nu            & 0.075 \\ \hline
\bar{b}\to\bar{c}\tau\nu           & 0.018 \\ \hline\hline
\sum\bar{b}\to\bar{c}              & 0.615 \\ \hline\hline
c\to s u \bar{d}                   & 0.905 \\ \hline
c\to s e \nu                       & 0.162 \\ \hline\hline
\sum c\to s                        & 1.229 \\ \hline\hline
\mbox{WA:}\,\bar{b} c\to c \bar{s} & 0.138 \\ \hline
\mbox{WA:}\,\bar{b} c\to \tau\nu   & 0.056 \\ \hline
\mbox{PI}                          &-0.124 \\ \hline\hline
\mbox{Total}                       & 1.914 \\ \hline\hline
\end{array}
$$
\caption{\label{table1}
Contributions to the $B_c$ meson decay rate in $
\mbox{ps}^{-1}$. The partial $\bar b$ and $c$ quark decay 
rates are understood to
include the nonperturbative corrections from dimension five operators
as discussed in the text. Cabibbo suppressed modes are implicitly
taken into account with the corresponding allowed channels. 
The input parameters are specified at the
beginning of Sect.~IV. WA=weak annihilation, PI=Pauli interference.}
\end{table}

\begin{table}[h]
\addtolength{\arraycolsep}{0.2cm}
$$
\begin{array}{|c||c|c|}
\hline
X & X_c & l_X \\ 
\hline\hline
m_b                & 5.0\,\mbox{GeV}     & -1.96 \\ \hline
m_c                & 1.5\,\mbox{GeV}     & -3.10 \\ \hline 
m_s                & 0.2\,\mbox{GeV}     &  0.14 \\ \hline
|V_{cb}|           & 0.04                & -0.72 \\ \hline
\mu_1              & 5.0\,\mbox{GeV}     &  0.02 \\ \hline
\mu_2              & 1.5\,\mbox{GeV}     &  0.13 \\ \hline
\mu_3              & 2.3\,\mbox{GeV}     & -0.07 \\ \hline
T                  & 0.37\,\mbox{GeV}    &  0.16 \\ \hline
M_{B_c^*}-M_{B_c}  & 0.073\,\mbox{GeV}   &  0.01 \\ \hline
f_{B_c}            & 0.5\,\mbox{GeV}     & -0.07 \\ \hline
M_{B_c}            & 6.26\,\mbox{GeV}    & -0.04 \\ \hline
\end{array}
$$
\caption{\label{table2}
Parameter dependence of $\tau_{B_c}$ against (small) variations of the 
parameter set $\{X_c\}$ specified at the beginning of Sect.~IV 
and repeated in the second column of the table. 
$l_X$ is defined by $\delta\tau_{B_c}(X_c)/\tau_{B_c}(X_c)\equiv 
l_X (\delta X/X_c)$, where $\delta\tau_{B_c}(X_c)=\tau_{B_c}(X_c+\delta 
X)-\tau_{B_c}(X_c)$. Thus, for instance, the entry for $X=m_c$ tells 
us that increasing $m_c$ from $1.5\,$GeV to $1.6\,$GeV decreases 
$\tau_{B_c}(X_c)=0.52\,$ps by $20.7\%$.
}
\end{table}

\newpage
\vspace*{2cm}

\end{document}